\documentclass[prb,twocolumn]{revtex4}
\usepackage[T1]{fontenc}
\usepackage[latin9]{inputenc}
\usepackage{prettyref}
\usepackage{textcomp}
\usepackage{amsmath}
\usepackage{graphicx}
\usepackage{amssymb}
\usepackage{esint}

\makeatletter

\newcommand{\lyxmathsym}[1]{\ifmmode\begingroup\def\b@ld{bold}
  \text{\ifx\math@version\b@ld\bfseries\fi#1}\endgroup\else#1\fi}

\usepackage{psfrag}

\makeatother

\usepackage[french]{babel}

\begin{document}

\title{Homogenization of nonlocal wire metamaterial via a renormalization
approach}

\author{A.I. C\u{a}buz$^{1}$, A. Nicolet$^{2}$, F. Zolla$^{2}$, D. Felbacq$^{3}$
and G. Bouchitté$^{4}$}

\affiliation {$^{1}$CNRS, Institut Fresnel, Campus de St. Jérôme, 13013 Marseille,
France\\
$^{2}$Université d'Aix-Marseille, Institut Fresnel, Campus de
St. Jérôme, 13013 Marseille, France\\
$^{3}$Université de Montpellier II, Groupe d'\'Etude des Semiconducteurs,
34095 Montpellier, France\\
$^{4}$Université de Toulon et du Var, BP 132, 83957 La Garde CEDEX,
France}

\begin{abstract}
It is well known that defining a local refractive index for a metamaterial
requires that the wavelength be large with respect to the scale of
its microscopic structure (generally the period). However, the converse
does not hold. There are simple structures, such as the infinite,
perfectly conducting wire medium, which remain non-local for arbitrarily
large wavelength-to-period ratios. In this work we extend these results
to the more realistic and relevant case of \emph{finite} wire media
with \emph{finite} conductivity. In the quasi-static regime the metamaterial
is described by a non-local permittivity which is obtained analytically
using a two-scale renormalization approach. Its accuracy is tested
and confirmed numerically via full vector 3D finite element calculations.
Moreover, finite wire media exhibit large absorption with small reflection,
while their low fill factor allows considerable freedom to control
other characteristics of the metamaterial such as its mechanical,
thermal or chemical robustness.
\end{abstract}

\maketitle

The effective medium theory of artificial metallo-dielectric structures
goes back to the beginning of the 20th century, with the work of Maxwell-Garnett
\cite{Maxwell-Garnett1904} and Wiener \cite{Wiener1912}. These,
and subsequent effective medium theories focused on disordered media
where only partial information on the microscopic structure was available.
A major step forward was made with the work of Kock, in the 1940s
\cite{Kock1948}. This time Lorentz theory \cite{Lorentz1902,Lorentz1916}
was used to \emph{design} artificial effective media, in a bottom
up fashion, as an array of scatterers. In the 1970s more mathematically
sophisticated methods emerged, where instead of seeking a limiting
effective \emph{medium} (equivalent in some suitably defined sense
to the structure of interest), one obtains a limiting \emph{equation
system} \cite{Bensoussan1978,Guenneau2000}, for the macroscopic electromagnetic
field in a given structure \cite{Felbacq1997,Bouchitte2004,Felbacq2005,Bouchitte2009}. 

In recent years, the advent of negative index metamaterials and composites
has led to increased interest in effective medium theories. The most
popular by far is of course the Lorentz theory approach, it being
the most accessible and intuitively appealing \cite{Elser2006}. However,
the usefulness of Lorentz theory is much diminished when one is interested
in materials where the size of objects is much larger than the distances
separating them, or materials which are strongly non-local, or in
which the scatterers are strongly coupled, leading to behavior of
a collective nature \cite{Cabuz2008,Cabuz2007a}. Contrary to common
intuition, non-local behavior persists, in certain structures, even
when the wavelength is much larger than the characteristic scale of
the structure; an excellent example is the wire medium studied by
Belov et al. \cite{Belov2002,Belov2003,Simovski2004a}. In these situations
the Lorentz model is no longer useful and more sophisticated techniques
are required.

\begin{figure*}

\begin{centering}
\includegraphics[scale=0.7]{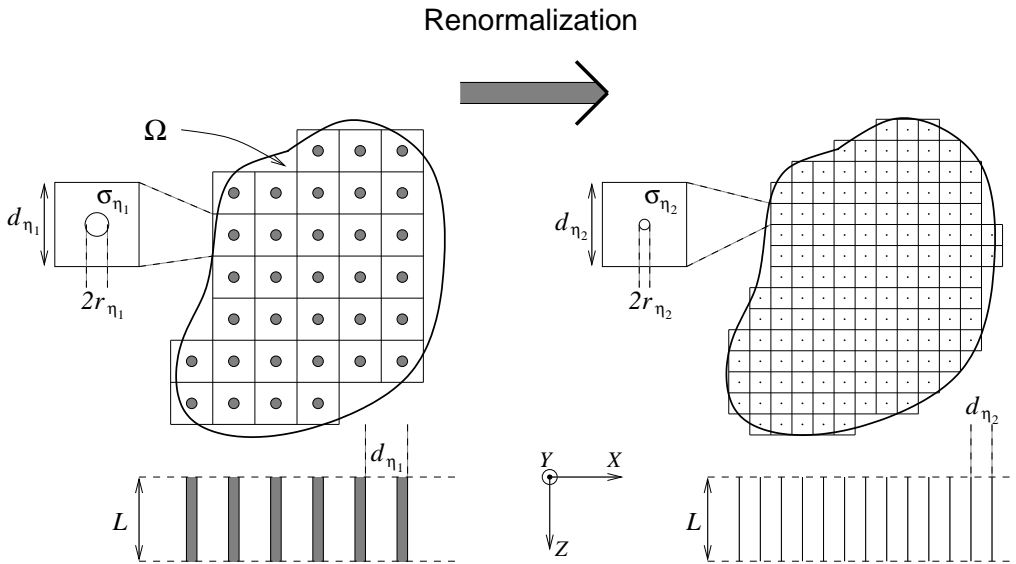}
\par\end{centering}

\caption{\label{fig:Structure-under-study}The bed-of-nails structure and the
renormalization process. The conducting fibers occupy a region $\Omega\subseteqq\mathbb{R}^{2}$,
are oriented in the $z$ direction, and the structure is periodic
in the $xy$ plane. Two renormalized structures are shown, corresponding
to $\eta_{1}$ and $\eta_{2}$ respectively, with $\eta_{1}>\eta_{2}$,
$d_{\eta_{1}}>d_{\eta_{2}}$, $\sigma_{\eta_{1}}<\sigma_{\eta_{2}}$
and $r_{\eta_{1}}/d_{\eta_{1}}>r_{\eta_{2}}/d_{\eta_{2}}$ (see Eqs.
\ref{eq:kappa} and \ref{eq:gamma}). The real physical structure
corresponds by definition to $\eta=1$: $d_{\eta=1}=d$. The length
$L$ and the wavelength $\lambda$ remain fixed, i.e., \emph{we are
homogenizing in the $xy$ plane only}. }

\end{figure*}

In this work we test and illustrate, for the first time, an effective
medium model of the finite conductivity finite wire medium (the {}``bed-of-nails''
structure, Fig. \ref{fig:Structure-under-study}) based on a two-scale
renormalization approach. Instead of letting the wavelength tend to
infinity, as customary in effective medium theories, we keep it fixed,
and let other geometrical parameters tend to zero. The advantage of
this approach is that it leaves us the possibility of keeping \emph{some}
of the geometrical parameters fixed (in this case the wire length
$L$), leading to a new type of \emph{partial} homogenization scheme.
To put it less formally, we would like to homogenize while keeping
the thickness fixed with respect to the wavelength, which prevents
us from letting $\lambda$ tend to infinity, so the only remaining
option is to make all the other dimensions (the wire radius $r$ and
the period $d$) tend to zero.

Unlike common practice in much of the metamaterials literature, we
include a detailed discussion of the model's domain of applicability,
so that an engineer may be able to quickly and efficiently decide
whether this kind of structure may be useful for a given purpose.

{\footnotesize }{\footnotesize \par}

The structure under study is a square biperiodic array of thin wires,
of length $L$, radius $r$ and conductivity $\sigma$. We note the
period $d$ and the wavelength $\lambda$. The renormalization (depicted
in Fig. \ref{fig:Structure-under-study}) involves a limiting process
whereby the three quantities: $r$, $d$ and $1/\sigma$ tend simultaneously
to zero. The parameter governing the limiting process is noted $\eta=d$,
the period. The asymptotics of the other two parameters, $\sigma$
and $r$, with respect to $\eta$ are described by \emph{fixed} parameters
$\kappa$ and $\gamma$ according to the following relations:\begin{eqnarray}
\kappa & = & \frac{\pi r_{\eta}^{2}\sigma_{\eta}}{\varepsilon_{0}\omega\eta^{2}}\label{eq:kappa}\\
\frac{1}{\gamma} & = & \eta^{2}\log(\frac{r_{\eta}}{\eta})\label{eq:gamma}\end{eqnarray}
where $\omega$ is the angular frequency of the electromagnetic field.
In other words the conductivity is renormalized inversely to the fill
factor $\theta_{\eta}=\frac{\pi r_{\eta}^{2}}{\eta^{2}}$, while the radius
is renormalized such that the expression $\eta^{2}\log(\frac{r_{\eta}}{\eta})$
remains constant. 

While these expressions may at first seem obscure, they have simple
intuitive interpretations. The first requires the current density
to remain constant during the renormalization. Notice that $\kappa$ is nothing other 
than the volume average of the imaginary part of the permittivity. Also, recall that the \emph{static}
admittance per unit length of a circular wire is given by \begin{eqnarray}
Y_{\mbox{wire}} & = & \pi r^{2}\sigma \label{eq:wireimpedance}\end{eqnarray}
and that the number of wires per unit area is given by $1/\eta^{2}$.
The second expression requires the average internal capacitance of
the wires to remain constant during renormalization. This feature
is known to be essential for their asymptotic behavior (see, for instance
Refs. \cite{Pendry1996,Pendry1998}). One may object that the expression
on the right side of Eq. \ref{eq:gamma} is valid for infinitely long
wires, whereas we are working with wires of finite length. Indeed,
as shown below, the model fails for short wires (comparable to the
period), but that configuration is best treated with the Lorentz approach
anyway \cite{Collin1991}, placing it outside our present scope.

The essential quantities in the rescaling process are therefore the
geometric quantities $r_{\eta}$, $\eta$, the material quantity $\sigma_{\eta}$
and the field quantities $E_{\eta}$ and $H_{\eta}.$ To these one
must also add a quantity characterizing the all important electric
field in the wires. This is noted $F_{\eta}$, it is non-zero only
\emph{inside} the wires, and is given by \begin{eqnarray*}
F_{\eta} & = & \frac{\kappa}{\theta_{\eta}}E_{\eta}\\
 & = & \frac{\sigma_{\eta}}{\varepsilon_{0}\omega}E_{\eta}.\end{eqnarray*}
$F_{\eta}$ has the units of electric field, and in the microscopic,
inhomogeneous picture it is clearly proportional to the current density.
In the macroscopic, homogeneous picture, however, it will correspond
to the polarization density $P$. More precisely we have $\lim_{\eta\rightarrow0}F_{\eta}=P/i\varepsilon_{0}$. 

The question to be answered now becomes: what happens in the limit
$\eta\rightarrow0$ ? The answer is that the fields converge (in a
precise sense described in Ref. \cite{Bouchitte2006}) to the unique
solution of the following system:\begin{equation}
\begin{cases}
\nabla\times E & =i\omega\mu_{0}H\\
\nabla\times H & =-i\omega\varepsilon_{0}(E+\frac{P}{\varepsilon_{0}}\hat{z})\\
\frac{\partial^{2}P_{z}}{\partial z^{2}}+\left(k_{0}^{2}+\frac{2i\pi\gamma}{\kappa}\right)P_{z} & =-2\pi\gamma\varepsilon_{0}E_{z},\, z\in[-L/2,L/2]\\
\frac{\partial P_{z}}{\partial z} & =0,\, z\in\{-L/2,L/2\}\end{cases}\label{eq:system1}\end{equation}
Before solving the system, let us first see what it tells us on a
more intuitive level. 

All field quantities above are effective, homogeneous quantities,
which have meaning when the wires have been replaced with a homogeneous
effective medium with an electric polarization density equal to $P$.
The equation which gives $P$ is an inhomogeneous Helmholtz equation
where the source term is given by the $z$ component of the electric
field $E_{z}$. The polarization satisfies Neumann conditions at the
upper and lower interfaces of the slab. It is not in general continuous
there because Maxwell's equations impose the continuity of the normal
component of the displacement field $D\equiv\varepsilon_{0}E+P$;
consequently, any jump in $E$ must be canceled by an equivalent jump
in $P/\varepsilon_{0}$. The dependence of $P$ on $E$, i.e., the
constitutive relation, takes the form of an integral. In this case
we are dealing with a one-dimensional inhomogeneous Helmholtz equation,
but this situation is slightly complicated by the fact that it is
valid on a bounded domain only (the thickness $L$ of the slab). The
polarization field has the form \begin{equation}
P(x,z_{0})=-2\pi\gamma\varepsilon_{0}\int_{-L/2}^{L/2}g(z,z_{0})E_{z}(x,z)dz\label{eq:nonlocal}\end{equation}
where $g(z,z_{0})$ is the Green function of the Helmholtz operator
on the bounded domain $z\in\left(-\frac{L}{2},\frac{L}{2}\right)$.
It takes the form (see Appendix A)\[
g(z,z_{0})=\frac{1}{K\sin(KL)}\cos\left[K(z_{<}+\frac{L}{2})\right]\cos\left[K(z_{>}-\frac{L}{2})\right]\]
where $K^{2}=k_{0}^{2}+\frac{2i\pi\gamma}{\kappa}$, $z_{<}=\min(z,z_{0})$
and $z_{>}=\max(z,z_{0})$. Relation \ref{eq:nonlocal} is clearly
a non-local constitutive relation because the value of the polarization
field at a position $z_{0}$ depends on values of the electric field
at positions different from $z_{0}$.

When the imaginary part of $K$ is large the integral above drops
off quickly. In the limit of small conductivity (and hence small $\kappa$),
the polarization becomes local for sufficiently large wavelengths.
In the opposite limit, for infinite conductivity and infinitely long
wires the integral covers all space (in the $z$ direction) and the
material is non-local, even in the long-wavelength regime. In fact
this can be seen immediately by doing a Fourier transform on the third
equation of system \ref{eq:system1} (with $\kappa\rightarrow\infty$): 

\[
\widehat{P_{z}}=\frac{-2\pi\gamma\varepsilon_{0}}{k_{z}^{2}-k_{0}^{2}}\widehat{E_{z}}\]
which gives \[
\varepsilon=1+\frac{2\pi\gamma}{k_{0}^{2}-k_{z}^{2}}\]
This is consistent with the findings of Belov et al. \cite{Belov2002,Belov2003,Simovski2004a}. 

Until now, the discussion has been independent of the actual shape
of the domain $\Omega$ (Fig. \ref{fig:Structure-under-study}). From
this point on, however, for purposes of illustration we specialize
to the case $\Omega=\mathbb{R}^{2}$, which is an infinite two dimensional
bed-of-nails, of thickness $L$, period $d$, wire radius $r$ and
conductivity $\sigma$. The effective medium is therefore a homogeneous
slab parallel to the $xy$ plane and of thickness $L$.

\begin{figure}
\begin{centering}
\includegraphics[scale=0.8]{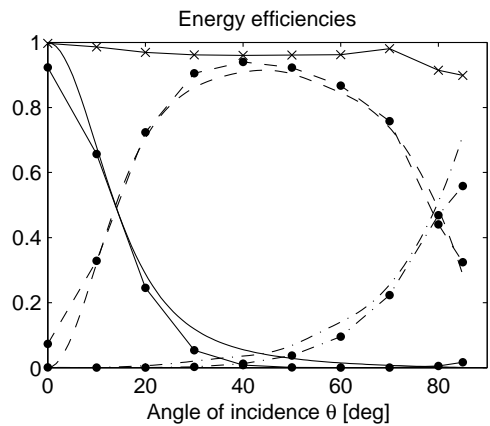}
\par\end{centering}

\caption{\label{fig:sweeptheta}Transmission (solid), reflection (dot-dashed)
and absorption (dashed) efficiency curves comparing the finite element
solution (dot markers) and the effective medium solution (no
markers) as a function of angle of incidence. The structure has a
conductivity $\sigma=8(\mbox{\ensuremath{\Omega}m})^{-1}$, period
$d=0.01\mbox{m}$, and dimensionless parameters $L/d=120$, $\lambda/d=20$,
$r/d=0.1$, and $\delta/d=4.6$. Computational constraints forced
us to use a very coarse mesh, which explains the approximate nature
of the energy conservation ($\times$ markers) of the finite element
model.}

\end{figure}

\begin{figure}
\begin{centering}
\includegraphics[scale=0.8]{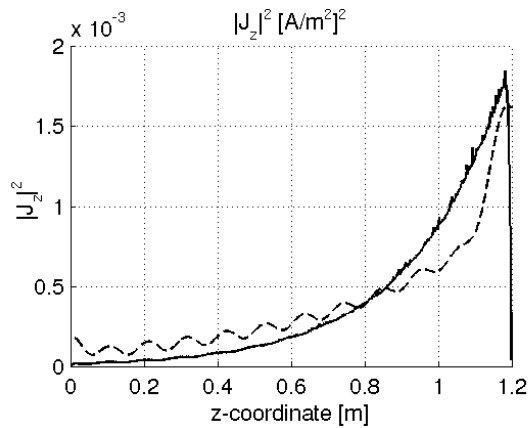}
\par\end{centering}

\caption{\label{fig:Current-density}Square of the current density for the
effective medium solution of Eqs. \ref{eq:systemsolve} (dashed)
and the finite element solution (solid) as a function of position
within the bed-of-nails structure (which is positioned in $z\in(0,L)$).
The structure is the same as in Fig. \ref{fig:sweeptheta}, illuminated
at an angle of incidence $\theta=40\lyxmathsym{\textdegree}$ from
the top.}

\end{figure}

\subsection*{Numerical results}

We now proceed to test the homogeneous model by comparing it with
3D full vector simulations of the structure, i.e. we must compare
the reflection, transmission and absorption coefficients and the current
distribution of the homogeneous problem with those of the original
bed-of-nails metamaterial. The solution to the homogeneous problem
is obtained by integrating system \ref{eq:system1} as described
in Appendix B.

The 3D full vector simulations of the bed-of-nails metamaterial were
done using the Comsol Multiphysics finite element method \cite{Dular1995}
software package. The periodicity was implemented using Floquet-Bloch
conditions \cite{Nicolet2004} in the two periodic directions ($x$
and $y$), and absorbing Perfectly Matched Layers \cite{Agha2008}
in the positive and negative $z$ directions. The linearity of the
materials in the structure was used to treat the incident field as
a localised source within the obstacle, as detailed in Ref. \cite{Demesy2007,Demesy2009a}.
The Comsol/Matlab scripts of the models used to produce the figures
below are available as online support material for the readers' convenience.

Figures \ref{fig:sweeptheta} and \ref{fig:Current-density} show
good agreement between the effective medium model and the finite element
simulation. Note that the current density behavior near the boundaries
differs between the effective medium model and the finite element
model. This is due to the fact that in the macroscopic, homogeneous
scenario, one speaks of a polarization field obeying Neumann boundary
conditions, as discussed above. In the microscopic scenario however,
we have a free conductor carrying current induced by an external electric
field. Since in our geometry at the given wavelength the capacitance
of the wire endpoints is very small, the accumulation of charge will
be correspondingly small, leading to an almost continuous normal component
of the electric field (and therefore also current). Numerically, it
seems as if the current goes to zero at the wire endpoints, even though
this is not strictly exact. Nevertheless, since in the homogeneous
limit the boundary condition of the current is of Neumann type, the
convergence of the renormalization process is clearly non-uniform
near the boundaries. This provides an additional explanation for requiring
long wires; we want the effect of the boundaries to be small. 

It must also be pointed out that the parameters of the particular
structure chosen for the illustration in Figs. \ref{fig:sweeptheta}
and \ref{fig:Current-density} were forced upon us by practical constraints:
finite element meshing of thin long circular wires requires very large
amounts of computer memory and time. Simulation of wires thinner than
$r/d=0.05$ is prohibitive. Consequently, in order to explore a wider
domain of the parameter space, we have taken advantage of the fact
that the structures we are interested in have $r\ll d$ and $\delta\gg r$.
Such thin conducting structures can be simulated much more efficiently
as lines of zero thickness \cite{Carpes2002} (i.e. \emph{edges},
in the finite element formulation) carying current and exhibiting
an equivalent \emph{linear impedance}. This approach gives excellent
results with a fraction of the computing power, and enables us to
model realistic structures that would otherwise be inaccessible. 

\begin{figure}
\begin{centering}
\includegraphics[scale=0.8]{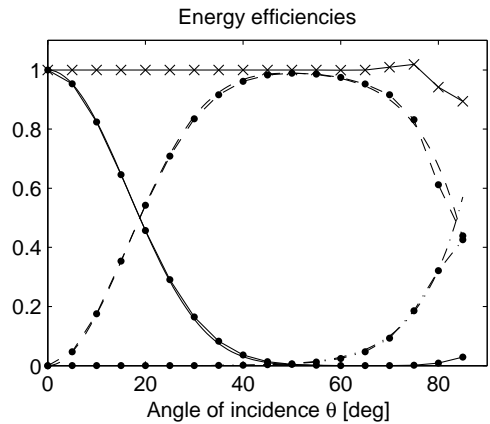}
\par\end{centering}

\caption{\label{fig:sweeptheta_smallr}Transmission (solid), reflection (dot-dashed)
and absorption (dashed) efficiency curves comparing the finite element
solution (dot markers) and the effective medium (no markers) as a
function of angle of incidence. The wire conductivity is that of Toray
T300\textsuperscript\textregistered ~carbon fibers $\sigma=5.89\cdot10^{4}(\mbox{\ensuremath{\Omega}m})^{-1}$.
The structure has period $d=0.01\mbox{m}$, and dimensionless parameters
$L/d=80$, $\lambda/d=20$, $r/d=3.5\cdot10^{-4}$, and $\delta/r=15$.
Energy conservation of the finite element model ($\times$ markers)
is respected to within better than one percent for most angles of
incidence. The departure around 80\textdegree{} is explained by the
poor performance of the PML absorbing layers when close to grazing
incidence.}

\end{figure}

\begin{figure}
\begin{centering}
\includegraphics[scale=0.8]{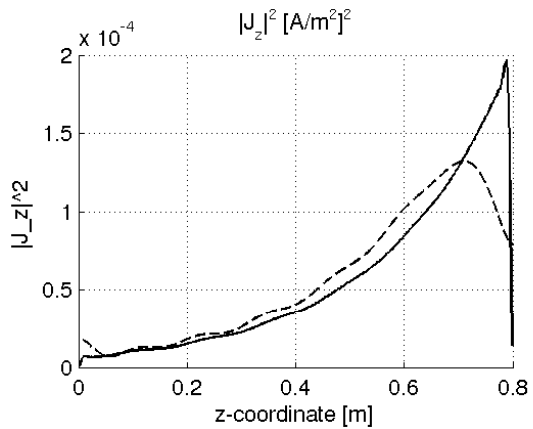}
\par\end{centering}

\caption{\label{fig:J_smallr}Square of the current density for the effective
medium model (dashed) and the finite element solution (solid) as a
function of position within the slab (which is positioned in $z\in(0,L)$).
The structure is the same as in Fig. \ref{fig:sweeptheta_smallr},
illuminated at an angle of incidence $\theta=40\lyxmathsym{\textdegree}$
from the top. Note that the surface areas under the two curves (in
this figure as well as Fig. \ref{fig:Current-density}) are the same
because they are proportional to the Joule dissipation rates, which
are seen to be equal from Fig. \ref{fig:sweeptheta_smallr} (and Fig.
\ref{fig:sweeptheta}) at the given angle of incidence.}

\end{figure}

For instance, Figs. \ref{fig:sweeptheta_smallr} and \ref{fig:J_smallr}
show the results of calculations for a structure of Toray T300\textsuperscript\textregistered
carbon fibers \cite{Toray} with a conductivity of: $\sigma=5.89\cdot10^{4}(\mbox{\ensuremath{\Omega}m})^{-1}$
and a radius of $3.5$ microns. The wires have an aspect ratio $L/r=2.28\times10^{5}$,
which is far beyond what would have been accessible by meshing the
interior of the wires. The finite element model of Fig. \ref{fig:sweeptheta}
(curves with markers), in which the interior of the wires is meshed,
is a problem with approx. 2.8 million degrees of freedom, which requires
at least 42 Gigabytes of available RAM to solve. By comparison, the
model of Fig. \ref{fig:sweeptheta_smallr} (curves with markers),
in which the wires are modeled as current carrying \emph{edges}, is
a problem of approx. 62 thousand degrees of freedom, which requires
less than one Gigabyte of available RAM and can therefore be solved
on any sufficiently recent desktop computer. 

\begin{figure}
\begin{centering}
\includegraphics[scale=0.8]{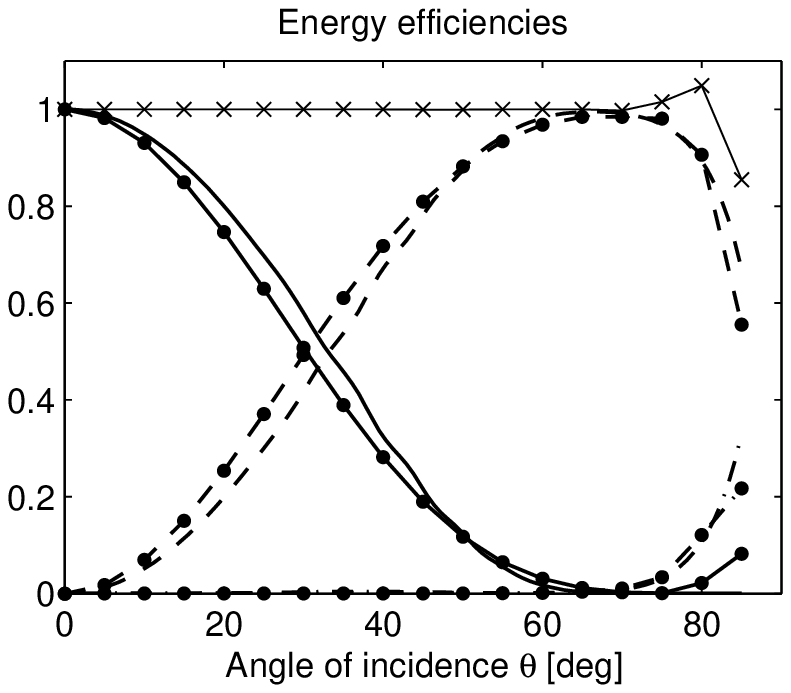}
\par\end{centering}

\caption{\label{fig:lowrefl}Transmission (solid), reflection (dot-dashed)
and absorption (dashed) efficiency curves comparing the finite element
solution (dot markers) and the effective medium (no markers) as a
function of angle of incidence. The structure has a conductivity $\sigma=1000(\mbox{\ensuremath{\Omega}m})^{-1}$,
period $d=0.01\mbox{m}$, and dimensionless parameters $L/d=50$,
$\lambda/d=8$, $r/d=0.002$, and $\delta/d=13$. The reflection remains
low for angles of incidence of up to 80\textdegree{} even as the Joule
absorption reaches almost 100\% for $\theta>60\lyxmathsym{\textdegree}$.
Energy conservation is indicated by the $\times$ markers.}

\end{figure}

\begin{figure}
\begin{centering}
\includegraphics[scale=0.8]{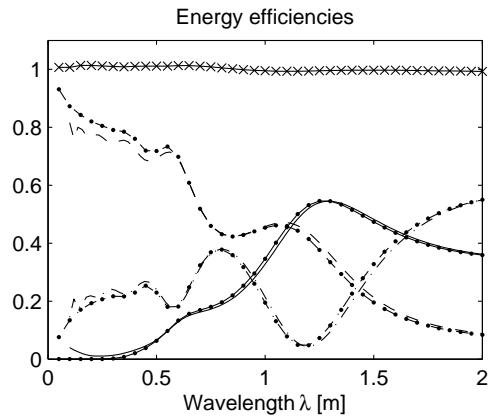}
\par\end{centering}

\caption{\label{fig:sweeptheta_smallr-1}Transmission (solid), reflection (dot-dashed)
and absorption (dashed) efficiency curves comparing the finite element
solution (dot markers) and the effective medium (no markers) as a
function of wavelength. Energy conservation for the finite element
model is labeled with $\times$ markers. The structure has a conductivity
$\sigma=3000(\mbox{\ensuremath{\Omega}m})^{-1}$ (in the semiconductor
domain), period $d=0.01\mbox{m}$, and dimensionless parameters $L/d=60$,
$r/d=0.003$, and the angle of incidence is $\theta=70\lyxmathsym{\textdegree}$.
$\delta/r$ runs approximately from $4$ to $25$ from left to right
over the domain of the plot. The model fails around $\lambda\lessapprox0.1\mbox{m}=10d$.}

\end{figure}

Figures \ref{fig:sweeptheta}, \ref{fig:Current-density}, \ref{fig:sweeptheta_smallr},
and \ref{fig:J_smallr} illustrate the behavior which is typical of
the model. The agreement remains good up to high incidence angles,
and over a large wavelength domain (Fig. \ref{fig:sweeptheta_smallr-1}).
The structure is transparent in normal incidence. For increasingly
oblique angles of incidence the absorption increases more or less
gradually, depending on the thickness $L$. The reflection is generally
low, though it increases when approaching grazing incidence. The low
reflection may be explained by the small radii of the wires: their
extremities have low capacitance, hence they exhibit very little charge
accumulation, leading to an almost continuous normal component of
the electric field. Certain configurations exhibit very low reflection
for almost all angles of incidence, see Fig.~\ref{fig:lowrefl} %
and Fig. \ref{fig:sweeptheta_smallr-1} around $\lambda=1.2\mbox{m}$.
The current density decreases roughly exponentially within the structure
due to absorption.

\subsection*{Domain of validity}

The boundaries of the domain of validity of the model are given by
four dimensionless parameters: the ratio of the skin depth to the
radius in the wires $\delta/r$, the ratio of the wire length to the
period $L/d$, the ratio of the wavelength to the period $\lambda/d$
and the ratio of the wire radius to the period $r/d$. 

The skin depth must be larger than the radius, due to the fact that
the impedance used in defining $\kappa$ (Eq. \ref{eq:wireimpedance})
is the \emph{static} impedance which differs from the \emph{quasistatic}
value by an imaginary inductive term $i\omega\mu/8\pi$ (see, for
instance, Ref. \cite{Ramo1994}). Requiring this term to be negligible
is equivalent to requiring that $\delta^{2}/r^{2}\gg1$. Moreover,
in the rescaling process the skindepth/radius ratio is given by \[
\frac{\delta_{\eta}}{r_{\eta}}=\frac{\lambda}{\eta}\sqrt{\frac{1}{2\pi\kappa}}.\]
Since $\eta$ approaches zero in the rescaling process, it is natural
to expect the homogeneous model to be valid when the skindepth is
large compared to the radius. 

In addition, recall that the definition of $\gamma$ in Eq. \ref{eq:gamma}
fixes the capacitance of the wires to the value for thin, long wires.
Consequently, we expect the model to hold for large $L/d$ and for
small $r/d$. To these, we must add the general requirement for all
effective medium models: the wavelength must be large compared to
the period.

Due to the large (four dimensional) parameter space, an \emph{exhaustive}
numerical exploration of the bed-of-nails structure is not feasible
in a reasonable timeframe. Still, our study has made it possible to
broadly determine the boundaries of the domain of applicability of
the effective medium model. Roughly, one must have $\lambda/d\gtrapprox7-12$,
$\delta/r\gtrapprox4-8$, $L/d\gtrapprox20-30$, $r/d\lessapprox10$.
Our (\emph{a fortiori} limited) numerical exploration of the parameter
space suggests that the skindepth-to-radius ratio is often the main
limiting factor, particularly when considering highly conducting wires.

\subsection*{Conclusion}

We have tested numerically the effective medium theory of the bed-of-nails
structure, whose rigorous mathematical foundation is described in
Ref. \cite{Bouchitte2006}. We have found good agreement between the
transmission, reflection and absorption efficiencies between the effective
medium model and a 3D finite element model, for a broad range of angles
of incidence and wavelengths. The current density in the real structure
corresponds to the polarization current density of the effective medium
model. The medium is nonlocal, meaning that the polarization field
depends on the electric field over a region of finite size. That dependence
is given by Eq. \ref{eq:nonlocal}. This nonlocal behavior also
means that the permittivity depends on the wavevector, so it can no
longer be seen, strictly, as a property of the \emph{medium}, but
rather, as a property of a given \emph{wave} propagating in the structure
\cite{Menzel2008,Rockstuhl2008}. 

The bed-of-nails structure is a medium exhibiting high absorption
with low reflection. It requires a very low filling fraction of conducting
material, but exhibits near perfect absorption over a wide range of
angles of incidence, for sufficiently large thicknesses. The low filling
fraction is useful because it allows the engineer to fill the space
between the wires with materials satisfying other design constraints,
such as mass density, or mechanical, chemical or thermal robustness.
The geometries studied here are transparent at normal incidence, but
this aspect can easily be rectified by slanting the wires by about
20\textdegree{} with respect to the upper and lower boundaries. This
design may therefore be used to obtain a near-perfect electromagnetic
absorber for all angles of incidence in a very straightforward way,
and with considerable freedom in the resulting mechanical, thermal
or chemical properties of the structure. We are currently exploring
more elaborate structures which may be modeled by the same scaling
technique: structures with thin wires in the x and/or y directions
as well as the z direction, or with wires curved helically, leading
to a non-trivial magnetic constitutive relation in addition to the
electric one.

\section*{Appendix A}

We require the Green function for the problem (see chapter II of Ref.
\cite{Schwartz2009}) \begin{eqnarray}
 & p^{''}+\alpha^{2}p=\beta E_{z}\nonumber \\
 & \mbox{with}\nonumber \\
 & p^{'}(-L/2)=p^{'}(L/2)=0.\label{eq:bndcond}\end{eqnarray}
For the purpose of this Appendix it is convenient to consider the
structure is positioned between$-L/2$ and $L/2$. The Green function
satisfies the equation\begin{eqnarray}
g^{''}+\alpha^{2}g=\delta_{z_{0}} & , & z_{0}\in\left(-\frac{L}{2},\frac{L}{2}\right)\label{eq:eqgreen}\end{eqnarray}
and may be written:\begin{eqnarray}
g(z,z_{0}) & = & Cu_{1}(z_{<})u_{2}(z_{>})\label{eq:green}\\
 & \mbox{with}\nonumber \\
z_{<} & = & \min(z,z_{0})\nonumber \\
z_{>} & = & \max(z,z_{0})\nonumber \end{eqnarray}
such that \begin{eqnarray*}
\mbox{when }z\in\left(-\frac{L}{2},z_{0}\right) & , & g=Cu_{1}(z)u_{2}(z_{0})\\
 & \mbox{and}\\
\mbox{when }z\in\left(z_{0},\frac{L}{2}\right) & , & g=Cu_{1}(z_{0})u_{2}(z)\end{eqnarray*}

Replacing form \ref{eq:green} into Eq. \ref{eq:eqgreen}
one obtains that $g$ must be continuous at $z_{0}$, its derivative
must have a jump discontinuity of 1, and the two functions $u_{1}$
and $u_{2}$ must be sinusoidal of wave constant $\alpha$:\begin{eqnarray*}
u_{1}(z) & = & A\cos(\alpha(z+L/2))\\
u_{2}(z) & = & B\cos(\alpha(z-L/2)).\end{eqnarray*}
By imposing the boundary conditions Eq. \ref{eq:bndcond} we
obtain \begin{eqnarray*}
u_{1}^{'}(-L/2) & = & 0\\
u_{2}^{'}(L/2) & = & 0\end{eqnarray*}
 and by requiring a jump discontinuity of 1 at $z_{0}$ we obtain
\[
ABC=\frac{1}{\alpha\sin(\alpha L)}\]
giving finally:\[
g(z,z_{0})=\frac{1}{\alpha\sin(\alpha L)}\cos(\alpha(z_{<}+L/2))\cos(\alpha(z_{>}-L/2))\]

\section*{Appendix B}

We now proceed to solve the homogeneous limit system~\ref{eq:system1}.
For convenience we position it in $z\in(0,L)$. Since we are dealing
with a system with translational invariance, a slab, we can split
the problem into two independent polarization cases: TE, where the
electric field is in the $xy$ plane, and TM, where the magnetic field
is in the $xy$ plane. However, since we are considering thin wires
(small volume fraction) the structure will be transparent to TE waves.
We therefore only have to consider TM waves. We choose a coordinate
system so that the plane of incidence is the $xz$ plane, with angle
of incidence $\theta$, in which case our unknowns will be $H_{y}$
and $P_{z}$. The translation invariance allows us to seek solutions
of the form:\begin{eqnarray*}
H_{y} & = & u(z)e^{i\alpha x}\\
P_{z} & = & p(z)e^{i\alpha x}\end{eqnarray*}
with: $\alpha=k_{0}\sin\theta$. Inserting these into system \ref{eq:system1}
we obtain a system of equations for $u$ and $p$:\begin{equation}
\begin{cases}
u^{\prime\prime}(z)+\left(k_{0}^{2}-\alpha^{2}\right)u(z) & =\alpha\omega p(z)\\
p^{\prime\prime}(z)+\left(k_{0}^{2}+\frac{2i\pi\gamma}{\kappa}-2\pi\gamma\right)p(z) & =\frac{2\pi\alpha\gamma}{\omega}u(z),\, z\in[0,L]\end{cases}\label{eq:systemsolve}\end{equation}

with the important boundary conditions: $p^{\prime}=0$ at $z=0$
and $z=L$, and $u$ and $u^{\prime}$ continuous everywhere. 

The objective is now to obtain the transfer matrix $T$ of the slab,
which relates the field $u$ and its derivative $u^{\prime}$ at the
bottom and the top of the slab:\begin{equation}
\left(\begin{array}{c}
u(L)\\
u^{\prime}(L)\end{array}\right)=T\left(\begin{array}{c}
u(0)\\
u^{\prime}(0)\end{array}\right).\label{eq:defT}\end{equation}
Once $T$ is known the reflection and transmission coefficients $r$
and $t$ can be obtained immediately from \begin{eqnarray}
r=e^{-2i\beta L}\frac{A+B}{A-B} & \mbox{and} & t=\frac{2e^{-i\beta L}}{A-B}\label{eq:refltrans}\\
A\equiv T_{11}-i\beta T_{12} & \mbox{and} & B\equiv\frac{T_{21}-i\beta T_{22}}{i\beta}\nonumber \end{eqnarray}
where $\beta=k_{0}\cos\theta=\sqrt{k_{0}^{2}-\alpha^{2}}$. 

We begin by integrating system \ref{eq:systemsolve}. Noting
$\delta^{2}=k_{0}^{2}+\frac{2i\pi\gamma}{\kappa}-2\pi\gamma$ for
readability, we rewrite the system as\begin{eqnarray}
W^{\prime\prime}(z) & = & -MW(z)\label{eq:systemsolvint}\end{eqnarray}
where \[
W(z)=\left(\begin{array}{c}
u(z)\\
p(z)\end{array}\right)\]
and \[
M=\left(\begin{array}{cc}
\beta^{2} & -\alpha\omega\\
-\frac{2\pi\alpha\gamma}{\omega} & \delta^{2}\end{array}\right).\]

The matrix $M$ can be diagonalized $M=QDQ^{-1}$ with $D=\mbox{diag}(K_{u}^{2},K_{p}^{2})$
so the system \ref{eq:systemsolvint} can be rewritten $Q^{-1}W^{\prime\prime}(z)=-DQ^{-1}W(z)$.
Since $Q$ is constant and known, this can be integrated directly,
and the general solution is then obtained as a sum of plane waves:\begin{eqnarray}
Q^{-1}W(z)=\left(\begin{array}{c}
A_{u}^{+}\exp(iK_{u}z)+A_{u}^{-}\exp(-iK_{u}z)\\
A_{p}^{+}\exp(iK_{p}z)+A_{p}^{-}\exp(-iK_{p}z)\end{array}\right)\label{eq:gensol}\end{eqnarray}
Once the integration performed, obtaining $T$ is now only a matter
of algebraic manipulation. $u$ and $p$ are now expressed in terms
of the elements of the matrix $Q$ and the coefficients $A_{u}^{+}$,
$A_{u}^{-}$, $A_{p}^{+}$ and $A_{p}^{-}$. However, recall that
we are not interested directly in these coefficients, but in the matrix
$T$. Since that matrix does not depend directly on $p$ the first
step is to eliminate the $A_{p}$s from the equation system. This
is done by making use of the boundary conditions. By differentiating
the bottom equation of system \ref{eq:gensol} we can obtain
$p^{\prime}$ as \begin{eqnarray*}
p^{\prime} & = & iK_{u}Q_{21}(A_{u}^{+}e^{iK_{u}z}-A_{u}^{-}e^{-iK_{u}z})\\
 &  & +iK_{p}Q_{22}(A_{p}^{+}e^{iK_{p}z}-A_{p}^{-}e^{-iK_{p}z}).\end{eqnarray*}
Setting this to zero at $z=0,L$ we can obtain the $A_{p}$s in terms
of the $A_{u}$s. Noting vectors \begin{eqnarray*}
\underline{A_{u}} & = & \left(\begin{array}{c}
A_{u}^{+}\\
A_{u}^{-}\end{array}\right)\\
\underline{A_{p}} & = & \left(\begin{array}{c}
A_{p}^{+}\\
A_{p}^{-}\end{array}\right),\end{eqnarray*}
we introduce the matrix \begin{eqnarray*}
C & = & -\frac{K_{u}Q_{21}}{K_{p}Q_{22}}\frac{1}{2i\sin(K_{p}L)}\\
 &  & \times\left(\begin{array}{cc}
e^{iK_{u}L}-e^{-iK_{p}L} & e^{-iK_{p}L}-e^{-iK_{u}L}\\
e^{iK_{u}L}-e^{iK_{p}L} & e^{iK_{p}L}-e^{-iK_{u}L}\end{array}\right)\end{eqnarray*}
so that \[
\underline{A_{p}}=C\underline{A_{u}}.\]
We are now in a position to express $W(z)$ in terms of $\underline{A_{u}}$
alone. Eq. \ref{eq:gensol} can be rewritten \begin{equation}
W(z)=QE(z)\underline{A_{u}}\label{eq:gensol_rewrite}\end{equation}
 where $E(z)$ is defined as \[
E(z)=\left(\begin{array}{cc}
e^{iK_{u}z} & e^{-iK_{u}z}\\
C_{11}e^{iK_{p}z}+C_{21}e^{-iK_{p}z} & C_{12}e^{iK_{p}z}+C_{22}e^{-iK_{p}z}\end{array}\right).\]
Eq. \ref{eq:gensol_rewrite} contains (within its first row)
the expression for $u$. But to obtain the transfer matrix $T$ we
also require $u^{\prime}$. We simply differentiate Eq. \ref{eq:gensol_rewrite}
to obtain \begin{equation}
W^{\prime}(z)=QE^{\prime}(z)\underline{A_{u}}.\label{eq:gensol_rewrite_dz}\end{equation}

By combining the first rows of Eqs. \ref{eq:gensol_rewrite}
and \ref{eq:gensol_rewrite_dz}, we are in a position to construct
the matrix $G(z)$ such that\[
\left(\begin{array}{c}
u(z)\\
u^{\prime}(z)\end{array}\right)=G(z)\underline{A_{u}}.\]
By writing this equation at $z=0$ and $z=L$ we obtain \[
\left(\begin{array}{c}
u(z)\\
u^{\prime}(z)\end{array}\right)=G(L)G(0)^{-1}\left(\begin{array}{c}
u(0)\\
u^{\prime}(0)\end{array}\right).\]
Comparing with Eq. \ref{eq:defT} we obtain the result we seek,
\[
T=G(L)G(0)^{-1},\]
leading to the reflection and transmission coefficients via Eqs. \ref{eq:refltrans}.
The Matlab script of the above manipulations is available as online
support material for the readers' convenience.

To summarize, we are now capable of modeling a structure with a given
$d$, $r$, $\sigma$, $L$ at a given incident field wavelength $\lambda$
in the following way. We first obtain the two rescaling parameters
$\kappa$ and $\gamma$ for the given structure using Eqs. \ref{eq:kappa}
and \ref{eq:gamma}. Then, we integrate system \ref{eq:systemsolve}
to obtain the reflection and transmission coefficients.

\bibliographystyle{apsrev}

\begin{thebibliography}{10}

\bibitem{Maxwell-Garnett1904}
J.~Maxwell-Garnett {\em Phil. Trans. R. Soc. Lond. A}, vol.~203, p.~385, 1904.

\bibitem{Wiener1912}
O.~Wiener {\em Abh. Math.-Phys. Konigl. Sachs. Ges.}, vol.~32, p.~509, 1912.

\bibitem{Kock1948}
W.~E. Kock, ``Metallic delay lenses,'' {\em Bell System Technical Journal},
  vol.~27, no.~1, pp.~58--82, 1948.

\bibitem{Lorentz1902}
H.~Lorentz {\em Proc. Roy. Acad., Amsterdam}, vol.~254, 1902.

\bibitem{Lorentz1916}
H.~Lorentz, {\em The theory of electrons and its applications to the phenomena
  of light and radiant heat}.
\newblock G.E. Stechert and Co., 1916.

\bibitem{Bensoussan1978}
A.~Bensoussan, J.~Lions, and G.~Papanicolaou, {\em Asymptotic Analysis for
  Periodic Structures}.
\newblock North-Holland, Amsterdam, 1978.

\bibitem{Guenneau2000}
S.~Guenneau and F.~Zolla, ``Homogenization of three-dimensional finite photonic
  crystals - abstract,'' {\em Journal of Electromagnetic Waves and
  Applications}, vol.~14, no.~4, pp.~529--530, 2000.

\bibitem{Felbacq1997}
D.~Felbacq and G.~Bouchitt\'{e}, ``Homogenization of a set of parallel fibers,''
{\em Waves in Random Media}, vol.~7, pp.~245, 1997.

\bibitem{Bouchitte2004}
G.~Bouchitt\'{e} and D.~Felbacq, ``Homogenization near resonances and
  artificial magnetism from dielectrics,'' {\em Comptes Rendus Mathematique},
  vol.~339, pp.~377--382, Sept. 2004.

\bibitem{Felbacq2005}
D.~Felbacq and G.~Bouchitt\'{e}, ``Theory of mesoscopic magnetism in photonic
  crystals,'' {\em Phys. Rev. Lett.}, vol.~94, no.~18, p.~183902, 2005.

\bibitem{Bouchitte2009}
G.~Bouchitt\'{e}, C.~Bourel, and D.~Felbacq, ``Homogenization of the {3D}
  {Maxwell} system near resonances and artificial magnetism,'' {\em Comptes
  Rendus de l'Academie des Sciences Serie I}, vol.~347, p.~571, 2009.

\bibitem{Elser2006}
J.~Elser, R.~Wangberg, V.~A. Podolskiy, and E.~E. Narimanov, ``Nanowire
  metamaterials with extreme optical anisotropy,'' {\em Applied Physics
  Letters}, vol.~89, 2006.

\bibitem{Cabuz2008}
A.~I. C\u{a}buz, D.~Felbacq, and D.~Cassagne, ``Spatial dispersion in
  negative-index composite metamaterials,'' {\em Phys. Rev. A}, vol.~77, no.~1,
  p.~013807, 2008.

\bibitem{Cabuz2007a}
A.~I. C\u{a}buz, {\em Electromagnetic metamaterials - From photonic crystals to
  negative index composites}.
\newblock PhD thesis, University of Montpellier II, 2007.

\bibitem{Belov2002}
P.~A. Belov, S.~A. Tretyakov, and A.~J. Viitanen, ``Dispersion and reflection
  properties of artificial media formed by regular lattices of ideally
  conducting wires,'' {\em Journal of Electromagnetic Waves and Applications},
  vol.~16, no.~8, pp.~1153--1170, 2002.

\bibitem{Belov2003}
P.~A. Belov, R.~Marques, S.~I. Maslovski, I.~S. Nefedov, M.~Silveirinha, C.~R.
  Simovski, and S.~A. Tretyakov, ``Strong spatial dispersion in wire media in
  the very large wavelength limit,'' {\em Phys. Rev. B}, vol.~67, no.~11,
  p.~113103, 2003.

\bibitem{Simovski2004a}
C.~R. Simovski and P.~A. Belov, ``Low-frequency spatial dispersion in wire
  media,'' {\em Physical Review E}, vol.~70, no.~4, p.~046616, 2004.

\bibitem{Pendry1996}
J.~B. Pendry, A.~J. Holden, W.~J. Stewart, and I.~Youngs, ``Extremely low
  frequency plasmons in metallic mesostructures,'' {\em Physical Review
  Letters}, vol.~76, no.~25, pp.~4773--4776, 1996.

\bibitem{Pendry1998}
J.~B. Pendry, A.~J. Holden, D.~J. Robbins, and W.~J. Stewart, ``Low frequency
  plasmons in thin-wire structures,'' {\em J. of Phys.-Cond. Matt.}, vol.~10,
  no.~22, p.~4785, 1998.

\bibitem{Collin1991}
R.~E. Collin, {\em Field theory of guided waves}.
\newblock IEEE Press, 1991.

\bibitem{Bouchitte2006}
G.~Bouchitt\'{e} and D.~Felbacq, ``Homogenization of a wire photonic crystal:
  The case of small volume fraction,'' {\em SIAM Journal On Applied
  Mathematics}, vol.~66, no.~6, pp.~2061--2084, 2006.

\bibitem{Dular1995}
P.~Dular, A.~Nicolet, A.~Genon, and W.~Legros, ``A discrete sequence associated
  with mixed finite-elements and its gauge condition for vector potentials,''
  {\em IEEE Transactions on Magnetics}, vol.~31, no.~3, pp.~1356--1359, 1995.
\bibitem{Nicolet2004}
A.~Nicolet, S.~Guenneau, C.~Geuzaine, and F.~Zolla, ``Modelling of
  electromagnetic waves in periodic media with finite elements,'' {\em Journal
  of Computational and Applied Mathematics}, vol.~168, no.~1-2, pp.~321--329,
  2004.

\bibitem{Agha2008}
Y.~{Ould Agha}, F.~Zolla, A.~Nicolet, and S.~Guenneau, ``On the use of pml for
  the computation of leaky modes - an application to microstructured optical
  fibres,'' {\em COMPEL -The International Journal for Computation and
  Mathematics in Electrical and Electronic Engineering}, vol.~27, no.~1,
  pp.~95--109, 2008.

\bibitem{Demesy2007}
G.~Dem\'{e}sy, F.~Zolla, A.~Nicolet, M.~Commandr\'{e}, and C.~Fossati, ``The
  finite element method as applied to the diffraction by an anisotropic
  grating,'' {\em Opt. Express}, vol.~15, no.~26, pp.~18089--18102, 2007.

\bibitem{Demesy2009a}
G.~Dem\'esy, F.~Zolla, A.~Nicolet, and M.~Commandre, ``Versatile full-vectorial
  finite element model for crossed gratings,'' {\em Optics Letters}, vol.~34,
  no.~14, pp.~2216--2218, 2009.

\bibitem{Carpes2002}
W.~{Carpes Jr.}, L.~Pichon, and A.~Razek, ``Analysis of the coupling of an
  incident wave with a wire inside a cavity using fem in frequency and time
  domains,'' {\em IEEE Transactions on Electromagnetic Compatibility}, vol.~44,
  p.~470, 2002.

\bibitem{Toray}
{Toray Carbon Fibers America Inc.}

\bibitem{Ramo1994}
S.~Ramo, J.~R. Whinnery, and T.~Van~Duzer, {\em Fields and Waves in
  Communication Electronics}.
\newblock John Wiley and Sons, third~ed., 1994.

\bibitem{Menzel2008}
C.~Menzel, C.~Rockstuhl, T.~Paul, F.~Lederer, and T.~Pertsch, ``Retrieving
  effective parameters for metamaterials at oblique incidence,'' {\em Phys.
  Rev. B}, vol.~77, pp.~195328--8, May 2008.

\bibitem{Rockstuhl2008}
C.~Rockstuhl, C.~Menzel, T.~Paul, T.~Pertsch, and F.~Lederer, ``Light
  propagation in a fishnet metamaterial,'' {\em Physical Review B}, vol.~78,
  p.~155102, Oct. 2008.

\bibitem{Schwartz2009}
L.~Schwartz, {\em Mathematics for the Physical Sciences}.
\newblock Dover Books on Mathematics, 2009.

\end{thebibliography}

\end{document}